\documentclass[10pt,twoside]{article}
\usepackage{pkas2}
\usepackage{graphicx}
\makehead{Goto}{LEGACY OF AKARI}

\def\lesssim{\mathrel{\hbox{\rlap{\hbox{\lower4pt\hbox{$\sim$}}}\hbox{$<$}}}}
\def\gtrsim{\mathrel{\hbox{\rlap{\hbox{\lower4pt\hbox{$\sim$}}}\hbox{$>$}}}}
\def\apj{ApJ}
\def\mnras{MNRAS}
\def\aap{A\&\hskip-1pt A}

\newcommand{\ltsima}{$\; \buildrel < \over \sim \;$}
\newcommand{\lsim}{\lower.5ex\hbox{\ltsima}}
\newcommand{\gtsima}{$\; \buildrel > \over \sim \;$}
\newcommand{\gsim}{\lower.5ex\hbox{\gtsima}}

\newcommand{\pasj}{{\it PASJ\ }}

\newcommand{\apjs}{{\it ApJS\ }}
\newcommand{\aj}{{\it AJ\ }}
\newcommand{\apjl}{{\it ApJL\ }}

\newcommand{\araa}{{\it ARAA\ }}

\begin{document}
\sloppy
\pagenumbering{arabic}
\twocolumn[
\pkastitle{27}{1}{2}{2012}
\begin{center}
{\large \bf {\sf
Cosmic star formation history and AGN evolution near and far: AKARI reveals both}}\vskip 0.5cm
{\sc Tomotsugu Goto, AKARI NEP team, and AKARI all sky survey team}\\
Dark Cosmology Centre, Niels Bohr Institute, University of Copenhagen\\
Juliane Maries Vej 30, 2100 Copenhagen 0, Denmark\\
{\it {E-mail: tomo@dark-cosmology.dk} }\\
\normalsize{\it (Received July 1, 2012; Accepted ????)}
\end{center}
\newabstract{
Understanding infrared (IR) luminosity is fundamental to understanding the cosmic star formation history and AGN evolution, since their most intense stages are often obscured by dust. Japanese infrared satellite, AKARI, provided unique data sets to probe this both at low and high redshifts.
The AKARI performed all sky survey in 6 IR bands (9, 18, 65, 90, 140, and 160$\mu$m) with 3-10 times better sensitivity than IRAS, covering the crucial far-IR wavelengths across the peak of the dust emission. Combined with a better spatial resolution, AKARI can much more precisely measure the total infrared luminosity ($L_{TIR}$) of individual galaxies, and thus, the total infrared luminosity density of the local Universe.
In the AKARI NEP deep field, we construct restframe 8$\mu$m, 12$\mu$m, and total infrared (TIR) luminosity functions (LFs) at 0.15$<z<$2.2 using 4128 infrared sources. A continuous filter coverage in the mid-IR wavelength (2.4, 3.2, 4.1, 7, 9, 11, 15, 18, and 24$\mu$m) by the AKARI satellite allows us to estimate restframe 8$\mu$m and 12$\mu$m luminosities without using a large extrapolation based on a SED fit, which was the largest uncertainty in previous work.
By combining these two results, we reveal dust-hidden cosmic star formation history and AGN evolution from $z$=0 to $z$=2.2, all probed by the AKARI satellite.
\vskip 0.5cm
{\em key words:} infrared - telescope: conferences - proceedings}
\vskip 0.15cm  \flushbottom
]

\newsection{INTRODUCTION}

Revealing the cosmic star formation history is one of the major goals of the observational astronomy. However, UV/optical estimation only provides us with a lower limit of the star formation rate (SFR) due to the obscuration by dust. 
A straightforward way to overcome this problem is to observe in the infrared, which can capture the star formation activity invisible in the UV. 
The superb sensitivities of recently launched Spitzer and AKARI satellites can revolutionize the field.

In the local Universe, often used IR LFs are from the IRAS (e.g., Sanders et al. 2003; Goto et al. 2011a) from 1980s, with only several hundred galaxies.
 In addition, bolometric infrared luminosities ($L_{IR,8-1000\mu m}$) of local galaxies were estimated using equation in P{\'e}rault (1987), which was a simple polynomial, obtained assuming a simple blackbody and dust emissivity. Furthermore, the reddest filter of IRAS was 100$\mu$m, which did not span the peak of the dust emission for most galaxies, leaving a great deal of uncertainty. 
Using deeper AKARI all sky survey data that cover up to 160$\mu$m,  we aim to measure local $L_{IR}$, and thereby the IR LF more accurately.

At higher redshifts, most of the Spitzer work relied on a large extrapolation from 24$\mu$m flux to estimate the 8, 12$\mu$m or total infrared (TIR) luminosity, due to the limited number of mid-IR filters.
AKARI has continuous filter coverage across the mid-IR wavelengths,  thus, allows us to estimate mid-IR luminosity without using a large $k$-correction based on the SED models, eliminating the largest uncertainty in previous work. 
By taking advantage of this, we present the restframe 8, 12$\mu$m TIR LFs, and thereby the cosmic star formation history derived from these using the AKARI NEP-Deep data.\begin{figure}
\begin{center}
\includegraphics[scale=0.6]{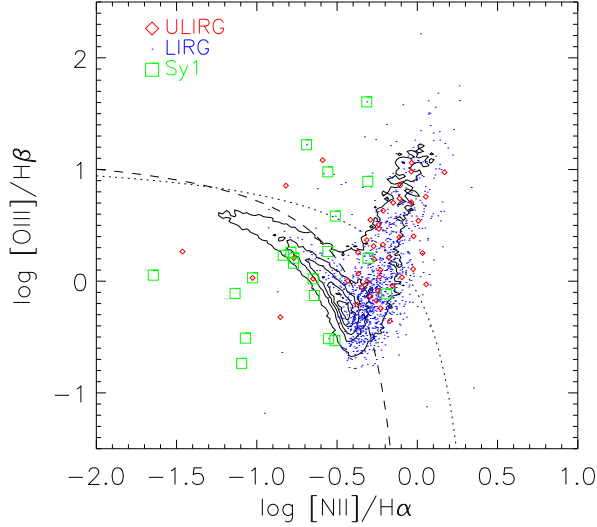}
\end{center}
\caption{
Emission line ratios used to select AGNs from the AKARI all sky sample. The contour shows distribution of all galaxies in the SDSS with $r<17.77$ (regardless of IR detection). 
The dotted line is the criterion between starbursts and AGNs described in Kewley et al. (2001).
The dashed line is the criterion by Kauffmann et al. (2003).
Galaxies with line ratios higher than the dotted line are regarded as AGNs. 
Galaxies below the dashed line are regarded as star-forming. 
Galaxies between the dashed and dotted lines are regarded as composites. 
The blue and red dots are for ULIRGs, LIRGs, respectively.
The green squares are Seyfert 1 galaxies identified by visual inspection of optical spectra. 
}\label{fig:BPT}
\end{figure}

\begin{figure}
\begin{center}
\includegraphics[scale=0.54]{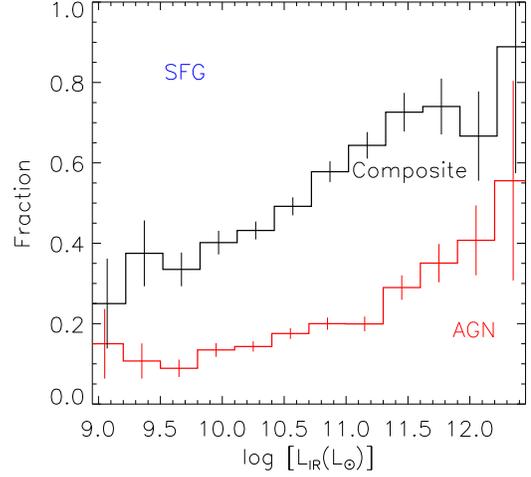}
\end{center}
\caption{Fractions of AGN and composite galaxies as  a function of $L_{IR}$.
AGN are classified using Kewley et al. (2001) among galaxies with all 4 lines measured.  Composite galaxies include those classified as AGN using Kauffmann et al. (2003).
}\label{fig:AGN_fractions}
\end{figure}

\begin{figure}[!ht]
\includegraphics[scale=0.54]{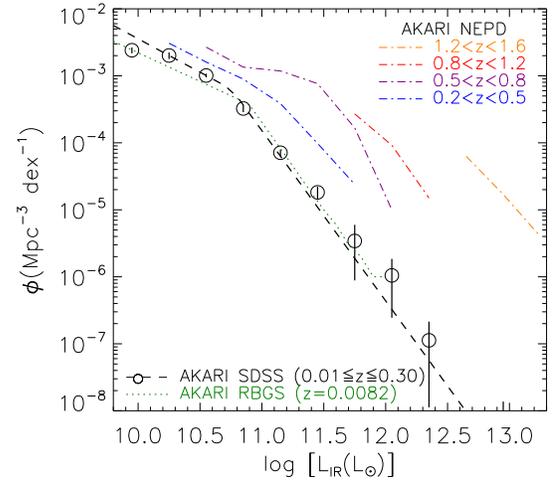}
\caption{
Infrared luminosity function of AKARI-SDSS galaxies. The $L_{IR}$ is measured using the AKARI 9, 18, 65, 90, 140 and 160$\mu$m fluxes through an SED fit. Errors are computed using 150 Monte Carlo simulations, added to a Poisson error.
The dotted lines show the best-fit double-power law. 
The green dotted lines show IR LF at $z$=0.0082 by Goto et al. (2011a).
The dashed-dotted lines are higher redshift results from the AKARI NEP deep field (Goto et al. 2010a,b).
}\label{fig:LF}
\end{figure}

\newsection{{\bfseries AKARI All Sky Survey: low-$z$ Universe}}

\begin{figure*}
\includegraphics[scale=0.35]{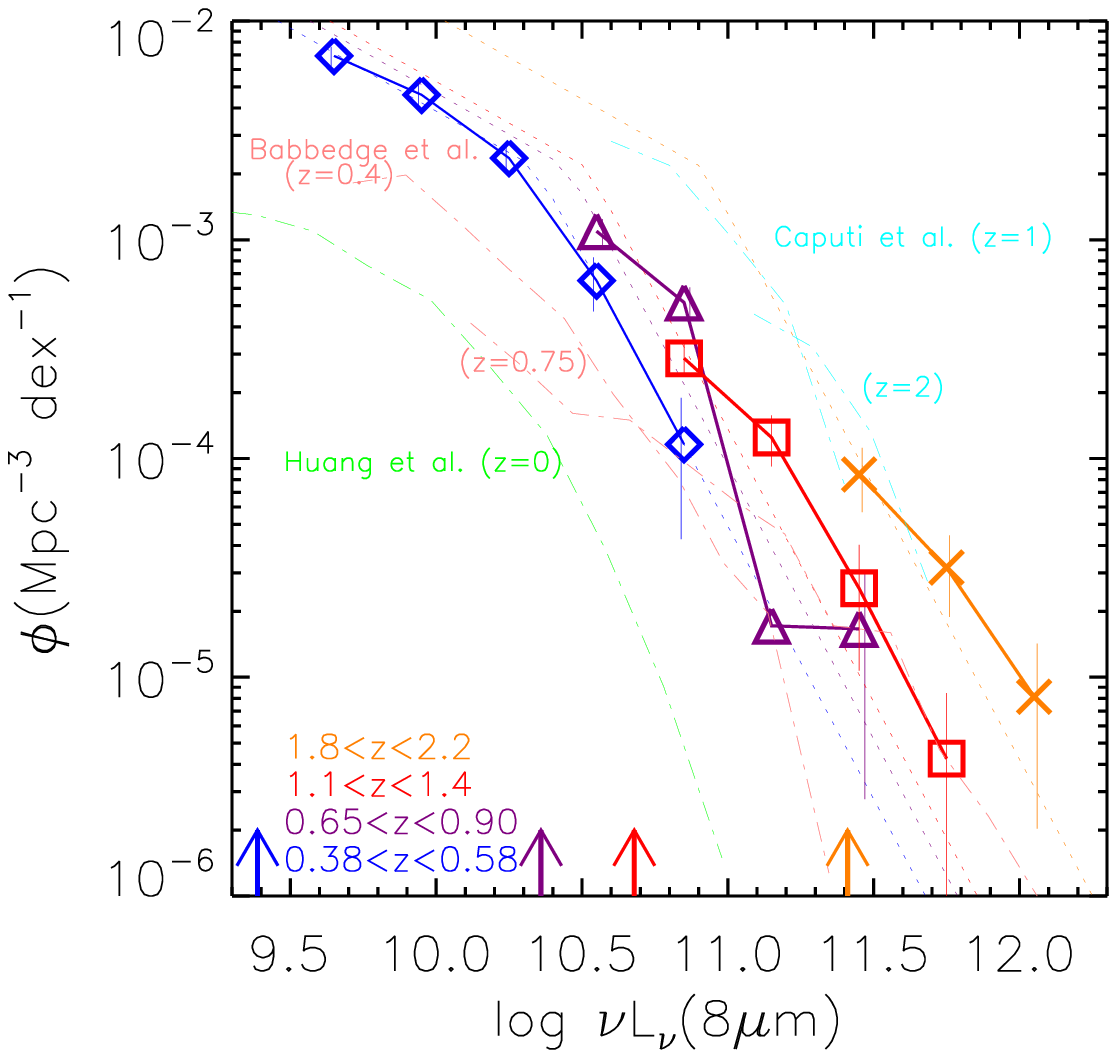}
\includegraphics[scale=0.35]{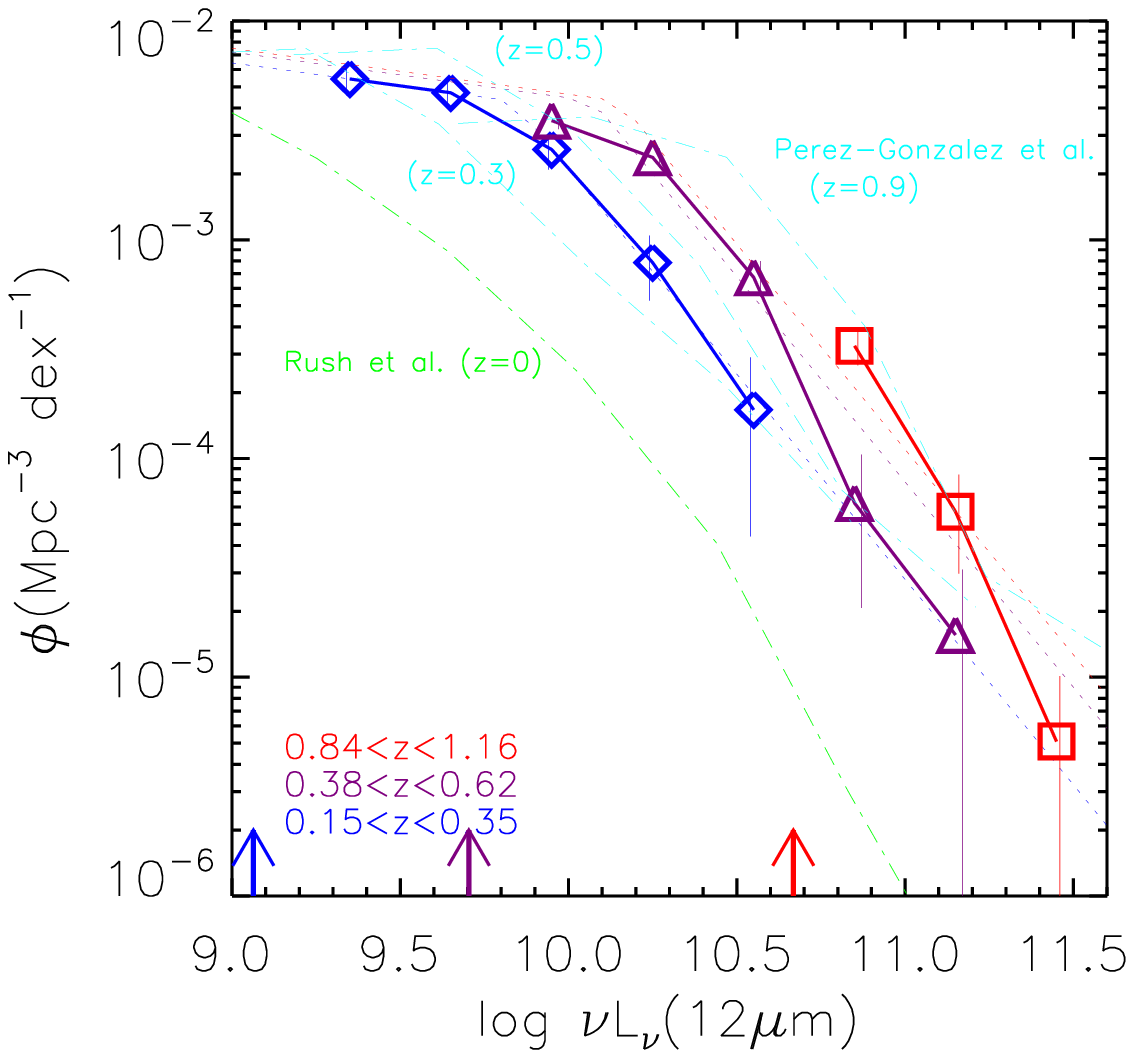}
\includegraphics[scale=0.35]{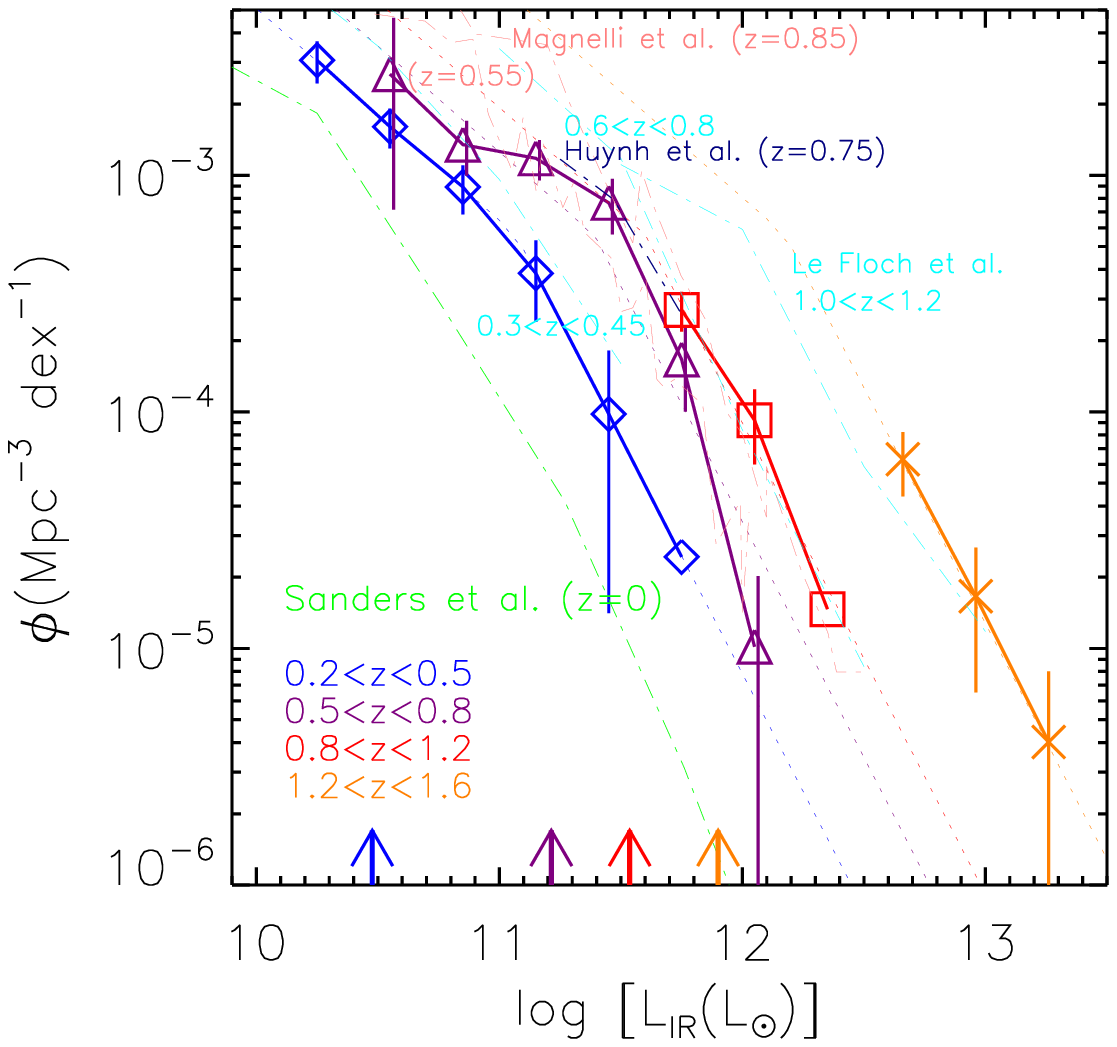}
\caption{
(left) Restframe  8$\mu$m LFs.
 The blue diamonds, purple triangles, red squares, and orange crosses show the 8$\mu$m LFs at $0.38<z<0.58, 0.65<z<0.90, 1.1<z<1.4$, and $1.8<z<2.2$, respectively. 
 The dotted lines show analytical fits with a double-power law.
 Vertical arrows show the 8$\mu$m luminosity corresponding to the flux limit at the central redshift in each redshift bin.
 Overplotted are Babbedge et al. (2006) in the pink dash-dotted lines, Caputi et al. (2007) in the cyan dash-dotted lines, and Huang et al. (2007) in the green dash-dotted lines. AGNs are excluded from the sample.
(middle)
Restframe  12$\mu$m LFs.
  The blue diamonds, purple triangles, and red squares show the 12$\mu$m LFs at $0.15<z<0.35, 0.38<z<0.62$, and $0.84<z<1.16$, respectively.
  Overplotted are P{\'e}rez-Gonz{\'a}lez et al. (2005) at $z$=0.3,0.5 and 0.9 in the cyan dash-dotted lines, and Rush et al. (1993) at $z$=0 in the green dash-dotted lines.
(right)
TIR LFs.  The redshift bins used are 0.2$<z<$0.5, 0.5$<z<$0.8, 0.8$<z<$1.2,  and 1.2$<z<$1.6. 
Overplotted are Le Floc'h et al. (2005), Magnelli et al. (2009), Huynh et al. (2007), and Sanders et al. (2003).
}\label{fig:8umlf}
\end{figure*}

\begin{figure*}
\begin{center}
 \includegraphics[height=.38\textheight]{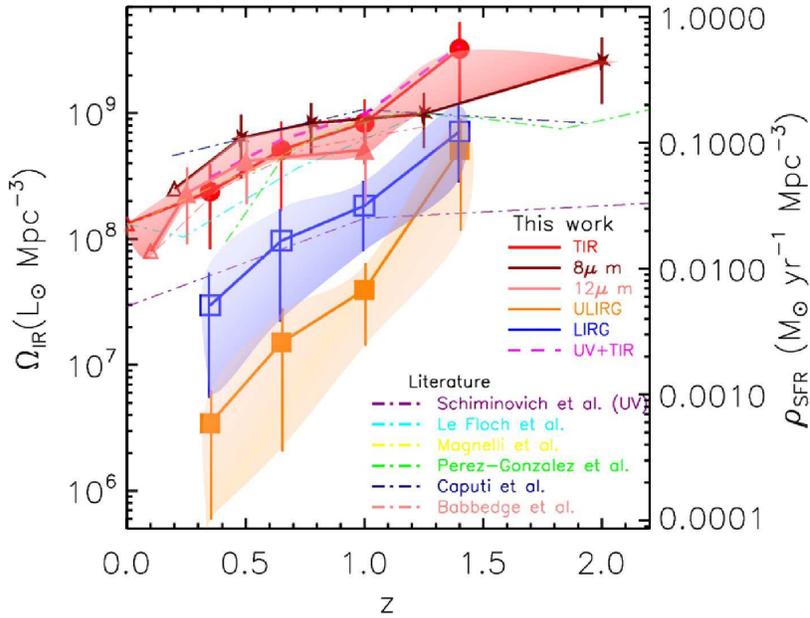}
\end{center}
\caption{
Evolution of TIR luminosity density based on TIR LFs (red circles), 8$\mu$m LFs (stars), and 12$\mu$m LFs (filled triangles). The blue open squares and orange filled squares  are for LIRG and ULIRGs only, also based on our $L_{TIR}$ LFs.
Overplotted dot-dashed lines are estimates from the literature: Le Floc'h et al. (2005), Magnelli et al. (2009), P{\'e}rez-Gonz{\'a}lez et al. (2005), Caputi et al. (2007),   and Babbedge et al. (2006) are in cyan, yellow, green, navy, and pink, respectively.
The purple dash-dotted line shows UV estimate by Schiminovich et al. (2005).
The pink dashed line shows the total estimate of IR (TIR LF) and UV.
}\label{fig:TLD_all}
\end{figure*}

 AKARI performed an all-sky survey in two mid-infrared bands (centered on 9 and 18 $\mu$m) and in four far-infrared bands (65, 90, 140, and 160$\mu$m). 
 In addition to the much improved sensitivity and spatial resolution over its precursor (the IRAS all-sky survey), the presence of 140 and 160$\mu$m bands is crucial to measure the peak of the dust emission in the FIR wavelength, and thus the $L_{IR}$ of galaxies.
We have cross-correlated the AKARI FIS bright source catalog with the SDSS DR7 galaxy catalog, obtaining 2357 spectroscopic redshifts.

\begin{figure}
\begin{center}
\includegraphics[scale=0.33]{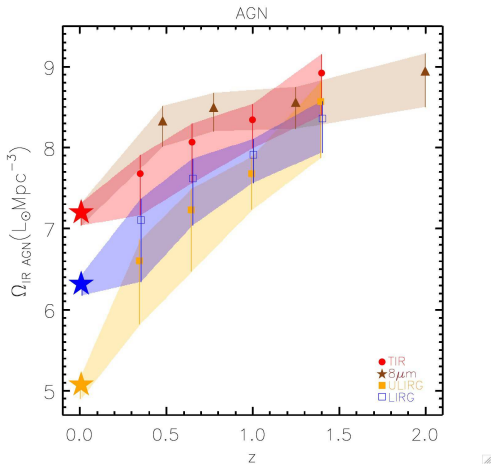}
\end{center}
\caption{
Evolution of TIR luminosity density by AGN.
 Results from the AKARI all sky survey is plotted with stars at $z$=0.0082. The red, blue and orange points show IR luminosity density from all AGN, from LIRG AGN only, and from ULIRG AGN only. 
Higher redshift results are from the AKARI NEP deep field (Goto et al. 2010a), with contribution from star forming galaxies removed.
 Brown triangles are $\Omega_{IR}^{AGN}$ computed from the 8$\mu$m LFs (Goto et al. 2010a).
}\label{fig:TLD_AGN_all}
\end{figure}

It is fundamental to separate IR contribution from two different physics; the star-formation and AGN activity.
In Fig. \ref{fig:BPT}, we use $[NII]/H\alpha$ vs $[OIII]/H\beta$ line ratios to classify galaxies into AGN or SFG (star-forming galaxies). It is interesting that majority of (U)LIRGs are aligned along the AGN branch of the diagram, implying the 
AGN fraction is high among (U)LIRGs. This is more clearly seen in Fig. \ref{fig:AGN_fractions}, where we plot fractions of AGN as a function of $L_{IR}$.
This results agree with previous AGN fraction estimates (Goto et al. 2005; Yuan et al. 2010).
 Improvement in this work is that due to much larger statistics, we were able to show fractions of AGN in much finer luminosity bins, more accurately quantifying the increase. Especially, a sudden increase of $f_{AGN}$ at log$L_{IR}>$11.3 is notable due to the increased statistics in this work.

For these galaxies, we estimated total IR luminosities ($L_{IR}$) by fitting the AKARI photometry with SED templates. 
We used the {\ttfamily  LePhare} code\footnote{http://www.cfht.hawaii.edu/$^{\sim}$arnouts/lephare.html} to fit the infrared part ($>$7$\mu$m) of the SED. 
We fit our AKARI FIR photometry with the SED templates from Chary \& Elbaz (2001; CHEL hereafter), which showed most promising results among SED models tested by Goto et al. (2011a).

With accurately measured $L_{IR}$, we are ready to construct IR LFs.
Since our sample is flux-limited at $r=17.7$ and $S_{90\mu m}=0.7Jy$, we need to correct for a volume effect to compute LFs.
We used the 1/$V_{\max}$ method. 
We estimated errors on the LFs with 150 Monte Carlo simulations, added to a Poisson error.

In Fig. \ref{fig:LF}, we show infrared LF of the AKARI-SDSS galaxies.
The median redshift of our sample galaxies is $z$=0.031.

Once we measured the LF,  we can estimate the total infrared luminosity density by integrating the LF, weighted by the luminosity. We used the best-fit double-power law to integrate outside the luminosity range in which we have data, to obtain estimates of the total infrared luminosity density, $\Omega_{IR}$. Note that outside of the luminosity range we have data ($L_{IR}>10^{12.5}L_{\odot}$ or $L_{IR}<10^{9.8}L_{\odot}$), the LFs are merely an extrapolation and thus uncertain.

The resulting total luminosity density is  $\Omega_{IR}$= (3.8$^{+5.8}_{-1.2})\times 10^{8}$ $L_{\odot}$Mpc$^{-3}$.
Errors are estimated by varying the fit within 1$\sigma$ of uncertainty in LFs.
 Out of  $\Omega_{IR}$, 1.1$\pm0.1$\% is produced by LIRG ($L_{IR}>10^{11}L_{\odot}$), and only 0.03$\pm$0.01\% is by ULIRG ($L_{IR}>10^{12}L_{\odot}$). Although these fractions are larger than $z$=0.0081 (Goto et al. 2011a), still a small fraction of  $\Omega_{IR}$ is produced by luminous infrared galaxies at $z$=0.031, in contrast to high-redshift Universe.

\begin{table}
 \centering
  \caption{Summary of evolution of infrared luminosity.
}\label{tab:omega}
  \begin{tabular}{@{}llrlllcccc@{}}
 \hline
$\Omega_{IR}^{SFG}$ & $\propto$(1+z)$^{4.1\pm0.4}$\\  
  $\Omega_{IR}^{SFG} (ULIRG)$&$\propto$(1+z)$^{10.0\pm0.5}$\\
  $\Omega_{IR}^{SFG}(LIRG)$&$\propto$(1+z)$^{6.5\pm0.5}$\\
$\Omega_{IR}^{AGN}$&$\propto$(1+z)$^{4.1\pm0.5}$\\
$\Omega_{IR}^{AGN}(ULIRG)$&$\propto$(1+z)$^{8.7\pm0.6}$\\
$\Omega_{IR}^{AGN}(LIRG)$&$\propto$(1+z)$^{5.4\pm0.5}$\\
 \hline
\end{tabular}
\end{table}

\newsection{{\bfseries AKARI NEP Deep Field: high-$z$ Universe}}

The AKARI has observed the NEP deep field (0.4 deg$^2$) in 9 filters ($N2,N3,N4,S7,S9W,S11,L15,L18W$ and $L24$) to the depths of 14.2, 11.0, 8.0, 48, 58, 71, 117, 121 and 275$\mu$Jy (5$\sigma$; Wada et al. 2008). 
This region is also observed in $BVRi'z'$ (Subaru),  $u'$ (CFHT), $FUV,NUV$ (GALEX), and $J,Ks$ (KPNO2m), with which we computed photo-z with $\frac{\Delta z}{1+z}$=0.043.  Objects which are better fit with a QSO template are removed from the analysis.
We used a total of 4128 IR sources down to 100 $\mu$Jy in the $L18$ filter.
We compute LFs using the 1/$V_{\max}$ method. Data are used to 5$\sigma$ with completeness correction. Errors of the LFs are from 1000 realization of Monte Carlo simulation.


\newsubsection{{\bfseries 8$\mu$m LF}}

Monochromatic 8$\mu$m luminosity ($L_{8\mu m}$) is known to correlate well with the TIR luminosity (Babbedge et al. 2006; Huang et al. 2007; Goto et al. 2011b), especially for star-forming galaxies because the rest-frame 8$\mu$m flux is dominated by prominent PAH features such as at 6.2, 7.7 and 8.6 $\mu$m.
 The left panel of Fig. \ref{fig:8umlf} shows a strong evolution of 8$\mu$m LFs.
 Overplotted previous work had to rely on SED models to estimate $L_{8\mu m}$ from the Spitzer $S_{24\mu m}$ in the MIR wavelengths where SED modeling is difficult due to the complicated PAH emissions. Here, AKARI's mid-IR bands are advantageous in  directly observing redshifted restframe 8$\mu$m flux in one of the AKARI's filters, leading to more reliable measurement of 8-$\mu$m LFs without uncertainty from the SED modeling. 

\newsubsection{{\bfseries 12$\mu$m LF}}

 12$\mu$m luminosity ($L_{12\mu m}$) represents mid-IR continuum, and known to correlate closely with TIR luminosity (P{\'e}rez-Gonz{\'a}lez et al. 2005). 
  The middle panel of Fig. \ref{fig:8umlf} shows a strong evoltuion of 12$\mu$m LFs.
 Here the agreement with previous work is better because (i)  12$\mu$m continuum is easier to be modeled, and (ii) the Spitzer also captures restframe 12$\mu$m in $S_{24\mu m}$ at $z$=1.

\newsubsection{{\bfseries TIR LF}}
 Lastly, we show the TIR LFs in the right panel of Fig. \ref{fig:8umlf}. 
We used Lagache et al. (2003)'s SED templates to fit the photometry using the AKARI bands at $>$6$\mu$m ($S7,S9W,S11,L15,L18W$ and $L24$). 
 The TIR LFs show a strong evolution compared to local LFs. 
 At $0.25<z<1.3$, $L^*_{TIR}$ evolves as $\propto (1+z)^{4.1\pm0.4}$.


\newsection{{\bfseries Cosmic star formation history}}

We fit LFs in Fig. \ref{fig:8umlf} with a double-power law, then integrate to estimate total infrared luminosity density at various $z$. The restframe 8 and 12$\mu$m LFs are converted to $L_{TIR}$ using P{\'e}rez-Gonz{\'a}lez et al. (2005) and Caputi et al. (2007) before integration.
 The resulting evolution of the TIR density is shown in Fig. \ref{fig:TLD_all}.
 The right axis shows the star formation density assuming Kennicutt (1998). 
 We obtain $\Omega_{IR}^{SFG}(z) \propto (1+z)^{4.1\pm 0.4}$.
Comparison to $\Omega_{UV}$ using Schiminovich et al. (2005) suggests that $\Omega_{TIR}$ explains 70\% of $\Omega_{total}$ at $z$=0.25, and that by $z$=1.3, 90\% of the cosmic SFD is explained by the infrared. This implies that $\Omega_{TIR}$ provides good approximation of the  $\Omega_{total}$ at $z>1$.
 
 In Fig. \ref{fig:TLD_all}, we also show the contributions to $\Omega_{TIR}$ from LIRGs and ULIRGs.
 From $z$=0.35 to $z$=1.4, $\Omega_{IR}$ by LIRGs increases by a factor of $\sim$1.6, and 
  $\Omega_{IR}$ by ULIRGs increases by a factor of $\sim$10.
More details are in Goto et al. (2010a).

\newsection{{\bfseries Cosmic AGN accreation history}}

We have separated the  $\Omega_{IR}^{SFG}$ from  $\Omega_{IR}^{AGN}$. 
We can also investigate $\Omega_{IR}^{AGN}$.

In Fig. \ref{fig:TLD_AGN_all}, we show the evolution of $\Omega_{IR}^{AGN}$, which shows a strong evolution with increasing redshift. 
 At a first glance, both $\Omega_{IR}^{AGN}$ and $\Omega_{IR}^{SFG}$ show rapid evolution, suggesting that the correlation between star formation and black hole accretion rate continues to hold at higher redshifts, i.e., galaxies and black holes seem to be evolving hand in hand.
 When we fit the evolution with (1+z)$^{\gamma}$, we found 
 $\Omega_{IR}^{AGN}\propto$(1+z)$^{4.1\pm0.5}$.
 A caveat, however, is that $\Omega_{IR}^{AGN}$ estimated in this work is likely to include IR emission from host galaxies of AGN, although in optical the AGN component dominates. Therefore, the final conclusion must be drawn from a multi-component fit based on better sampling in FIR by Herschel or SPICA, to separate AGN/SFG contribution to $L_{IR}$.
The contribution by ULIRGs quickly increases toward higher redshift;  By $z$=1.5, it exceeds that from LIRGs. Indeed, we found 
$\Omega_{IR}^{AGN}(ULIRG)\propto$(1+z)$^{8.7\pm0.6}$ and 
$\Omega_{IR}^{AGN}(LIRG)\propto$(1+z)$^{5.4\pm0.5}$.
Summary of the evolution of $\Omega_{IR}$ is presentedtj in Table \ref{tab:omega} (See Goto et al. 2011b for more details).

\end{description}

\end{document}